\documentclass[mathleft]{an}  
\usepackage{graphicx}
\usepackage{times}
\usepackage{longtable}
\usepackage{natbib}
\bibpunct{(}{)}{;}{a}{}{,} 
\def\msun{M$_\odot$}

\def\css{CSS081231:071126$+$440405}
\def\nick{J071126}

\begin{document}

\Pagespan{000}{000}
\Yearpublication{2014}%
\Yearsubmission{2014}%
\Month{00}%   
\Volume{000}%  
\Issue{00}% 

\title{Multi-epoch time-resolved photometry of the eclipsing polar \css}

\author{A.D.~Schwope\inst{1}\fnmsep\thanks{Corresponding author:
  \email{aschwope@aip.de}\newline}
\and
F.~Mackebrandt\inst{1}
\and
B.D.~Thinius\inst{2}
\and
C.~Littlefield\inst{3,4}
\and
P.~Garnavich\inst{3}
\and
A.~Oksanen\inst{5}
\and
T.~Granzer\inst{1}
}
\authorrunning{A.D.~Schwope et al.}
\institute{Leibniz-Institut f\"ur Astrophysik Potsdam (AIP),
              An der Sternwarte 16, 14482 Potsdam, Germany
              \and
Inastars Observatory, Hermann-Struve-Str. 10, 14469 Potsdam, Germany
\and
Department of Physics, University of Notre Dame, 225 Nieuwland Science Hall, Notre Dame, IN 46556, USA
\and
Department of Astronomy, Wesleyan University, Middletown, CT 06459, USA
\and
Hankasalmi Observatory,
Jyv\"askyl\"an Sirius ry, Vertaalantie 419, 40270 Palokka, Finland 
}

\received{December 4, 2014}
\accepted{January 21, 2015}
\publonline{}

\abstract{The eclipsing polar \css\ turned bright ($V_{\rm max}\sim 14.5$) 
in late 2008 and was subsequently observed intensively with small and medium-sized 
telescopes. A homogeneous analysis of this comprehensive dataset comprising 109 
eclipse epochs is presented and a linear ephemeris covering the five 
years of observations, about 24000 orbital cycles, is derived. 
Formally this sets rather tight constraints on the mass of a hypothetical circumbinary planet, 
$M_{\rm pl} \leq 2$\,M$_{\rm Jup}$. This preliminary result needs consolidation by long-term 
monitoring of the source.  The eclipse lasts $433.08 \pm 0.65$\,s, and 
the orbital inclination is found to be $i=79.3\degr - 83.7\degr$. The centre of the bright phase
displays accretion-rate dependent azimuthal shifts. No accretion geometry is found that 
explains all observational constraints, suggesting a complex accretion geometry with possible
pole switches and a likely non-dipolar field geometry.
}


\maketitle
%
%________________________________________________________________

\section{Introduction}
Jupiter-sized planets on Jupiter-like wide orbits were discovered in recent years
around eclipsing close binary stars with white-dwarf primaries and main-sequence
secondary stars (WDMS binaries). Such discoveries were possible via the light-travel time effect, 
i.e.~via periodic shifts of the eclipses of the central binary, and they require precise 
eclipse timing and long monitoring campaigns. Suitable target stars are relatively rare, to date
less than 100 such objects are known.
Circumbinary planets were found orbiting accreting and non-accreting binaries of WDMS type, 
i.e.~cataclysmic binaries (CVs) and pre-CVs \citep[see e.g.][]{beuermann+10,qian+11}.

Orbital period variations in those objects may come from different sources, they may be caused
by activity cycles of the secondary stars or by circumbinary planets. In addition, magnetic braking 
of the binary and gravitational radiation will occur but their strength is not  
enough to make an observable impact on the time-scale of a scientific career. 
Some objects show strong, quasi-regular variability in their eclipse
arrival times that cannot be explained by current models \citep[see the claimed discovery 
of a planetary system around HU Aqr; ][]{schwarz+09,qian+11,horner+11, wittenmyer+12,
gozdz+12,schwope+thinius14,bours+14}.

\css\ (henceforth \nick)
was discovered as variable transient-like object in the Catalina Real-Time Transient 
Survey \citep{drake+09} when
it suddenly turned bright from $V \sim 18$ to $V\sim14.5$ in the fall of the year 2008. Following 
vsnet-alert \#10867 Denisenko \& Korotkiy (2008, reported in CVnet discussion \#1208) 
obtained the first time-resolved photometric observations. 
They confirmed a maximum brightness of 14.4 and reported the occurrence of deep eclipses, 
which were used to determine an orbital period of 116.6\,m. The shape of the light-curve 
led them to classify the object as an eclipsing polar. In its high state it was found to be similarly bright as
the template eclipsing polar HU Aquarii.

\begin{table*}[t]
\label{t:facs}
\caption{Facilities used in this study and their observational set-up}
\begin{tabular}{lrccrrrr}	
Observatory & Aperture & Detector & Filter & $T_{\rm exp}$ & \# frames & latency & syst. unc. \\
 & (cm) & & &  (s) & & (s) & (s) \\
\hline
HO & 40 & SBIG STL-1001E & clear &  30 & 6730 & 0.0 & 3.00\\ 
Nonndorf & 25.4 & Videocam.~WAT-120N & --  & 2 & 10218 & 0.0 & 2.56\\
VATT 2009& 180 &  VATT4K & B/V & 20,30/30 & 426,321/608 & 0.0 & 0.0\\ 
VATT 2010& 180 & VATT4K & U/B & 30/30 & 526/517 & 0.0 & 0.0\\
VATT 2010& 180 & VATT4K & V & 10,15,30 & 112,549,206 & 0.0 & 0.0\\
VATT 2013& 180 & VATT4K & R/vilS$^\ast$ & 10/20 & 306/222 & 0.0 & 0.0 \\
STELLA 2012 & 120 & CCD & g & 15 & 543 & 3.0 & 0.1 \\
STELLA 2013 & 120 & CCD & g & 60 & 724 & 3.0 & 0.1 \\
IOP & 36 & SBIG ST-8 3 & Astrometrik CLS & 3 & 6845 & 0.0 & 0.0 \\
SCT28 & 28 & SBIG ST-8 3 & clear & 8,10,15,30 & 6304,1353,64,438 &1.5 & 0.0\\
SLKT & 80 & SBIG STL-1001 3 & Luminance & 2 & 222 & 2.6 & 0.0\\
\hline  
\end{tabular}
$^\ast$ the Vilnius S filter has maximum transparency at H$\alpha$ and a FWHM of 180\,\AA.
\end{table*}

Polars are magnetic cataclysmic variable stars. Contrary to normal CVs the strong magnetic moment 
of the accreting white dwarf keeps both stars in synchronous rotation. The magnetic field suppresses 
the formation of an accretion disk. Instead the accreted matter is guided towards the magnetic pole(s)
where the accretion luminosity eventually is released, part of it as strongly polarized cyclotron radiation.  
Polars are sources of soft and hard X-rays that are originating from regions close to the magnetic poles. 
The absence of an accretion disk results in short response times to changes of the mass transfer rate
that may occur on time scales from seconds to years. During low states, when the accretion stream 
is reduced to a trickle, the optical light is dominated by the photospheres of the stellar constituents
plus perhaps some cyclotron radiation from residual accretion. During high accretion states 
the spectral energy distribution of the binary is dominated by accretion-induced radiation
which outshines the photospheric radiation.

Bright eclipsing objects like HU Aqr and now \nick\ are prime targets for detailed investigations 
addressing the accretion physics and are qualified for high-time resolution observations 
to determine precise eclipse times.

The first printed publications on \nick\  are by \cite{thorne+10} and \cite{katysheva+shugarov12} 
who were refining the eclipse ephemeris and discussing the shape of the light-curve in low, intermediate
and high states of accretion. The object occasionally shows a pronounced pre-eclipse dip 
due to the intervening accretion stream or accretion curtain, whose location in phase in principle 
helps to uncover the accretion geometry. 

At its maximum brightness the variable star is within the reach of small telescopes 
and was targeted by amateurs at several sites. Amateur equipment may allow short 
integration times and their results can be used to study the LTT-effect. Lower signal-to-noise 
per image compared to professional telescopes can 
be compensated by 'infinite' observation time available at private observatories.

Motivated by recent reports of circumbinary planets, by the large brightness of \nick, and by the 
availability of a large body of archival data made available via the internet, we 
decided to obtain further time-resolved photometric observations of \nick\ and 
analyze the archival and new own data in a homogeneous way to derive a reliable 
ephemeris for the centre of the eclipse and to search for possible systematic deviations
that could give a hint for the occurrence of an unseen companion. 

In Sect.~\ref{s:obs} we describe both new and archival observations and how the data
were treated to determine eclipse times for the individual epochs. 
In Sect.~\ref{s:eph} we derive an updated eclipse ephemeris, in Sect.~\ref{s:lcs} 
we analyse properties of the optical light curves and derive 
binary parameters and discuss the possible accretion geometry in Sect.~\ref{s:con}.

%----------------------------------------------------------- 
\begin{figure}[t]
\resizebox{\hsize}{!}{\includegraphics[clip=]{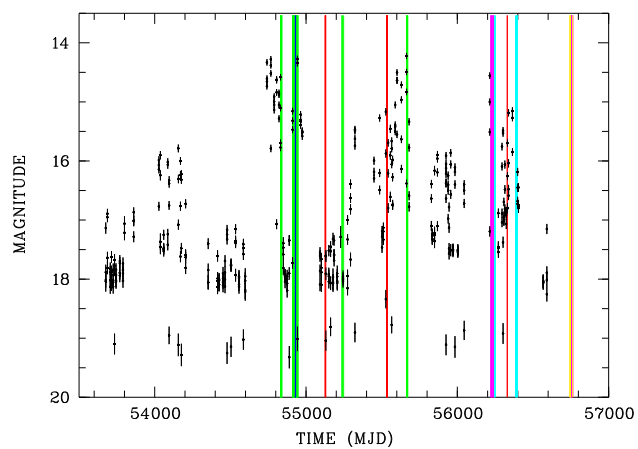}}
\caption{Long-term light curve of  \css\ derived from the CSS-database. Vertical lines
indicate the epochs used here for the determination of the binary period. Color indicates 
the equipment used (green -- HO, blue - Dangl,  red -- VATT, cyan -- STELLA, 
magenta -- Notre Dame (SCT28 \& SLKT), yellow -- IOP)
\label{f:poa}}
\end{figure}
%

%__________________________________________________________________

\section{Observations and data reduction}
\label{s:obs} 
In this paper new and archival observations of \nick\ are reported. New observations 
were obtained with the 1.2m STELLA/WiFSIP telescope, from the Inastars Observatory Potsdam (IOP), 
with the 1.8m Vatican Advanced Technology Telescope (VATT), with the newly inaugurated 80cm 
Sarah L.~Krizmanich telescope (SLKT) and with 
a 28\,cm Schmidt-Cassegrain telescope (shorthand SCT28), both owned by the 
University of Notre Dame (Indiana).
Archival observations were obtained with the VATT in 2009, 
by the Hankasalmi observatory (HO, Finland) and by Gerhard Dangl via his 
website\footnote{\tt http://www.dangl.at}. 
The VATT data from 2009 were originally published by \citet{thorne+10} but re-analyzed here 
in the same manner as all other data for the sake of homogeneity.
The HO-data were retrieved from the webpage of the observatory\footnote{\tt http://murtoinen.dynds.org}. 
The whole data set comprises time-resolved photometric observations obtained during more than
50 nights at the various locations. 
Instead of listing every single night of observation that was used in this study we indicate 
each night for which we could derive an eclipse epoch in the CSS long-term light curve 
displayed in Fig.~\ref{f:poa}.
Nevertheless, an overview of the facilities used and the main observational parameters 
is given in Table \ref{t:facs}.

\begin{figure}[t]
\resizebox{\hsize}{!}{\includegraphics[clip=]{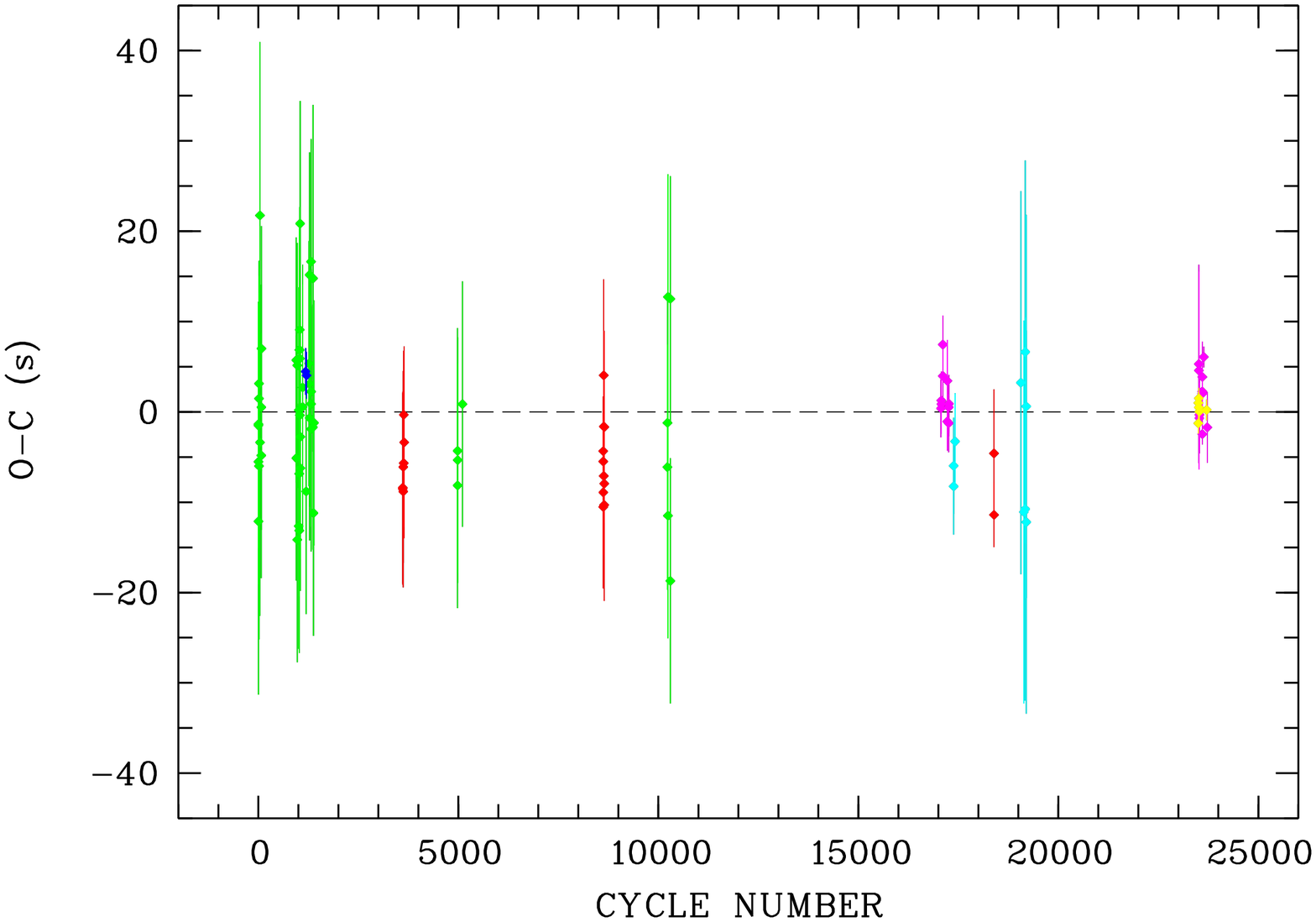}}
\caption{Observed minus calculated times of eclipse centres of \nick.
Colors are the same as in Fig.~\ref{f:poa}
}
\label{f:omc}
\end{figure}

\nick\ was put on the observing program of the Hankasalmi Observatory after its transient
nature was reported and was intensively observed during 2009 and occasionally in 2010 and 2011.
All those data were obtained with an integration time of 30\,s. 

The observations by Dangl from Nonndorf (Austria) were recorded with a video system 
(for details see his webpage) and binned afterwards to a time resolution of 2 seconds. 
A GPS-clock was directly connected to the video-grabber. The timing 
accuracy per frame is 2.56\,s (Dangl, private communication).
 
Details of the observations with the VATT in 2009 were described by \citet{thorne+10} and the 
description will not be repeated here. Additional data in UBV bands were obtained during 
three nights in 2010 and in the R and Vilnius S bands in 2013. 

%=== crtsfolded
\begin{figure}[t]
\resizebox{\hsize}{!}{\includegraphics[clip=]{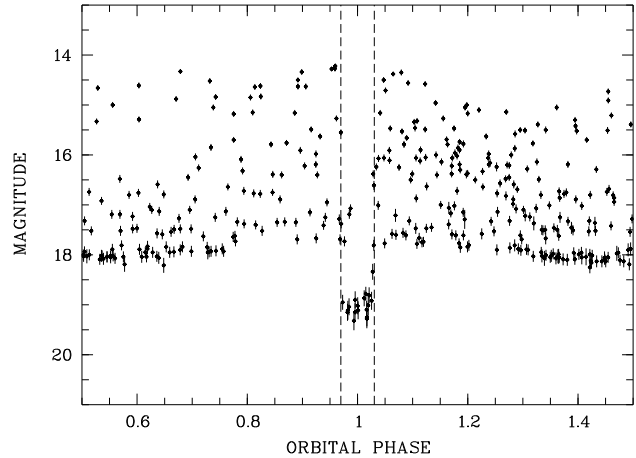}}
\caption{CRTS data folded over binary phase}
\label{f:crtsfol}
\end{figure}
%=== end figure

Observations with an 28\,cm Schmidt-Casegrain telescope (SCT28) owned by the University 
of Notre Dame were performed during 
eight  nights in 2012 and four nights in 2014, mostly in white light but occasionally during two 
nights in 2012 through B and R filters, respectively. 

The 80cm SLKT, also owned by the University of Notre Dame,  was used during  two commissioning nights 
of the telescope for un-filtered observations with integration times as short as 2 seconds.

\nick\ was observed from Inastars observatory Potsdam during four nights 
in March and April 2014. The instrument and the observational setup was the same 
as described in \cite{schwope+thinius14}.
  
The robotic telescope STELLA-1 \citep[aperture 1.2 m, ][]{strassmeier+04} is equipped with the wide-field
imaging photometer WiFSIP and was used during two nights in November 2012 and during 
five nights in April 2013 for time-resolved photometric observations of \nick. All CCD-frames
were taken through an SDSS g-filter. 

The raw CCD-frames obtained from the various sites were bias-corrected, dark-subtracted and flat-fielded using 
calibration data obtained for this purpose. In the context of the current study the timing 
of individual CCD-frames is of utmost importance. The anticipated start of individual exposures 
is stored in the FITS-headers in the DATE-OBS keyword. However, CCD pre-flashes 
and other delays due to the operating system may give rise to a delayed shutter opening. 
Whenever possible, this shutter latency 
was determined and added to the measured times of eclipse ingress and egress. When 
this was not possible, a systematic error of any timing value was taken into account.
Without measuring the shutter latency the widely used MaximDL CCD-software 
does not write decimal seconds to the FITS-headers but cuts time to the integer second. 
When integer seconds were written by the CCD software, half a second was added 
to the measured times and a systematic uncertainty of 0.5 seconds was added
to the error budget. The measured or estimated shutter latencies together 
with the systematic timing uncertainties are listed in columns (7) and (8) of Tab.~\ref{t:facs}. 

\begin{figure}[t]
\resizebox{\hsize}{!}{\includegraphics[clip=]{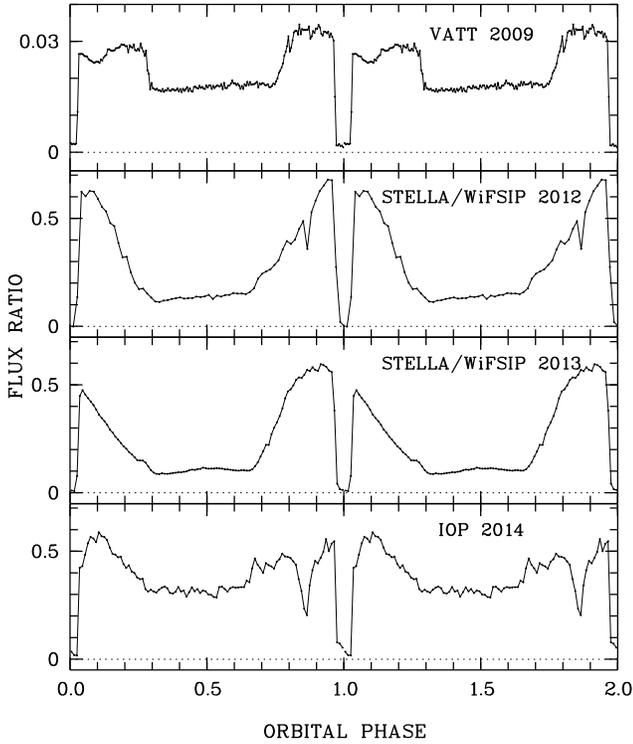}}
\caption{Phase-averaged light curves of \nick\ obtained between 2010 
and 2014 from various sites. All data are shown twice for clarity and were phased using 
the ephemeris of Eq.~\ref{eq1} and averaged into 100 phase bins 
(66 phase bins for the 2012 observations with STELLA).
Note the different scale used in the first panel.}
\label{f:lcfol}
\end{figure}

Afterwards, differential photometry using ESO-MIDAS routines was performed
with respect to comparison stars \#139 or \#142 on the AAVSO star chart 
 (\#139: RA(J2000) = 7:11:13.17, DE(J2000) = 44:03:59.7, $B=14.696, V=13.918$; 
 \#142: RA(J2000) = 7:11:35.50, DE(J2000) = 44:04:03.3, $B=15.167, V=14.158$).
The default choice was to use star \#139. The flux ratios and differential magnitudes 
shown in Figs.~\ref{f:lcfol} and \ref{f:ecl} refer to that star.

% Oksanen data 
%----------------------------------------------------------- 
\begin{figure*}[t]
\resizebox{\hsize}{!}{\includegraphics[clip=]{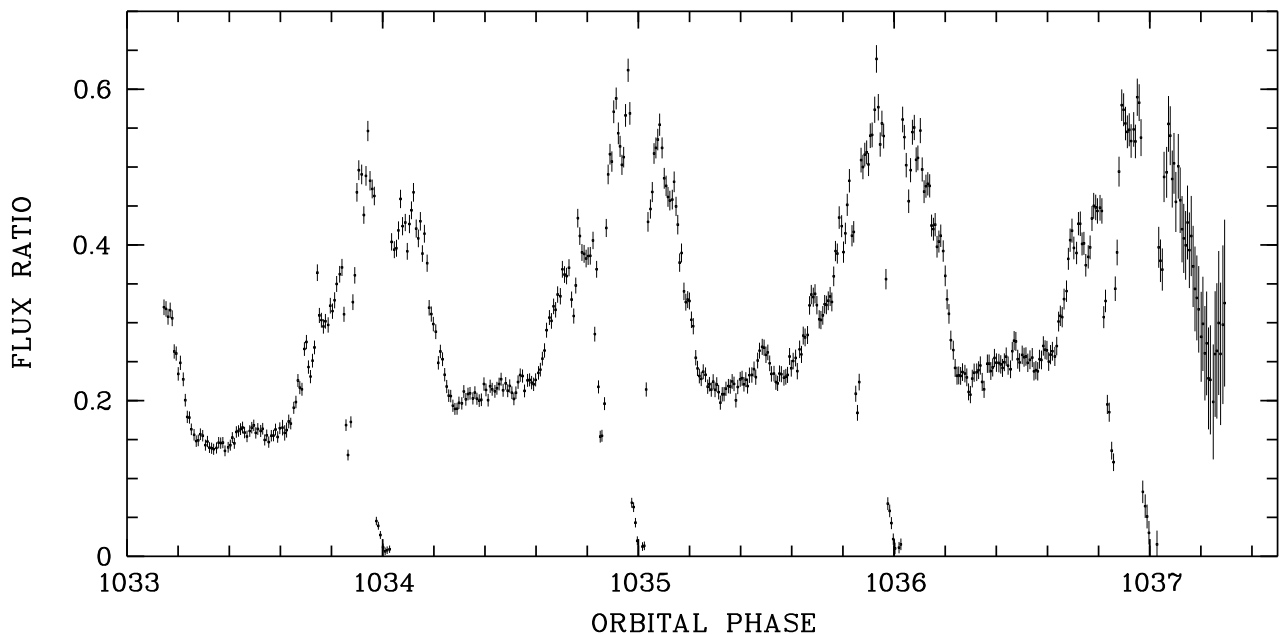}}
\caption{Time-resolved photometry obtained on March 26, 2009,
from Hankasalmi observatory. Shown is the flux ratio between the 
target and comparison \#139 from the AAVSO chart for the target.}
\label{f:lcoks1}
\end{figure*}

The centroids of the stellar images of both the target and the comparison star
were determined via two-dimensional Gaussian fits on a number of images 
with clean detections of both stars. The positions were used to determine  
the positional offset in pixel units between the two stars. 
Then aperture photometry was performed on each CCD frame using concentric apertures for 
the object (plus sky background) and the sky background, 
centered initially on the comparison star, then at the offset position and always using the 
same apertures for target and comparison star.
With this procedure a brightness measurement was 
possible even during eclipses of \nick, when the source became too faint 
for centering. Errors of individual measurements
were determined taken into account the sky brightness, the object brightness 
and the read-noise properties of the CCD. The flux ratio 
between the target and the comparison star was further investigated
for timing of the eclipse and for a discussion of the shape of the light curve(s).

%----------------------------------------------------------- 
\begin{figure}[t]
\resizebox{\hsize}{!}{\includegraphics[clip=]{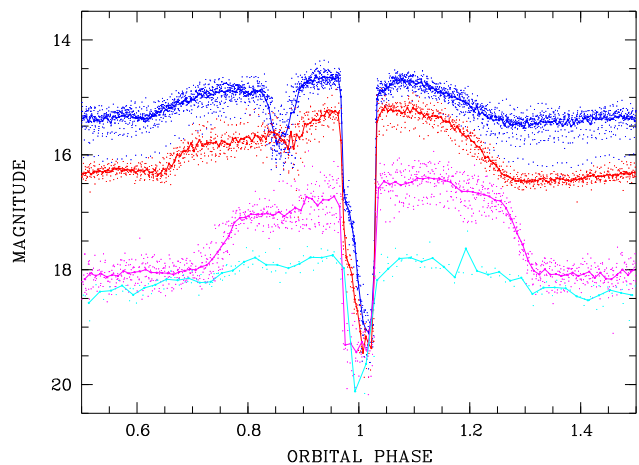}}
\caption{Phase-averaged light curves of \nick\ obtained in 2009
from Hankasalmi observatory through high, intermediate, low and very low states, respectively. 
The data were averaged into 200, 200, 100, and 50 phase
bins from top to bottom using the ephemeris of Eq.~\ref{eq1}
(high state: March 26, 27, and April 1, 8, 14, 15, $22-25$; 
intermediate state: January 1, 2, 4, and March 19; 
low state: January 7, March 21, 24, and 25;
very low state: January 16, September 19, 20).
}
\label{f:lcoks2}
\end{figure}

%----------------------------------------------------------- 

\section{Analysis and results}
\subsection{A precise long-term ephemeris \label{s:eph}}
The times of all CCD frames were corrected to refer to mid-exposure, 
corrected for arrival at the solar system barycenter
and converted to barycentric dynamical time (TDB) which equals terrestrial time (TT) 
at our precision.

The times of individual ingress and egress were measured 
by averaging a few data 
points before and after ingress/egress, computing the half-light intensity and reading 
the times with a cursor from a graph of the light curve 
\citep[see ][ for a graphical representation]{schwope+thinius14}. 
Uncertainties of individual measurements were set to half the integration time 
of individual CCD frames, unless a data gap degrades the timing accuracy further. 
There were occasions when only one eclipse feature, either the ingress or the egress, 
could be determined. In order to include those eclipses in the analysis as well
a two-stage process was chosen to derive timings of the eclipse centre.
Firstly the mean eclipse length was determined using all fully covered eclipses.
A weighted average of 102 individual measurements gave $\Delta t_{\rm ecl} = 
433.08 \pm 0.65$\,s, which corresponds to $\Delta \phi = 0.061596 \pm 0.000092$ phase units. 
Then the eclipse centers were determined from the timings of the ingress and/or egress
corrected for half the eclipse length. 
This procedure resulted in one (ingress or egress) or two (ingress and egress) independent  
measurements of the eclipse centre. In the latter case the finally accepted value was the weighted 
average of the two measurements.
All eclipse timings are listed in Tab.~\ref{t:ecl}. 

A weighted linear regression to all 109 data points yields the linear ephemeris for the eclipse centre 
\begin{eqnarray}
{\rm BJD(TDB)} =  2454833.207868(14) + \\ E \times 0.0813768094(6) \nonumber
\label{eq1}
\end{eqnarray}
(numbers in parenthesis give formal 1$\sigma$ uncertainties, 
reduced $\chi_\nu^2=0.92$ for 107 d.o.f.). The residuals with respect 
to this linear fit are shown in Fig.~\ref{f:omc}. The diagram and the 
fitting statistics show that all data are in agreement with the linear fit. 

%=== figure eclipse shape and depth

\begin{figure}[t]
\resizebox{\hsize}{!}{\includegraphics[clip=]{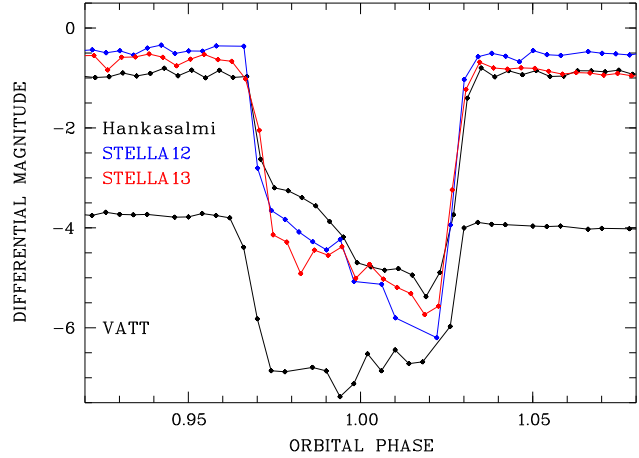}}
\caption{Shape of the eclipse in high and low states. }
\label{f:ecl}
\end{figure}
%=== end figure
%=== crtseclispe
\begin{figure}[h]
\resizebox{\hsize}{!}{\includegraphics[clip=]{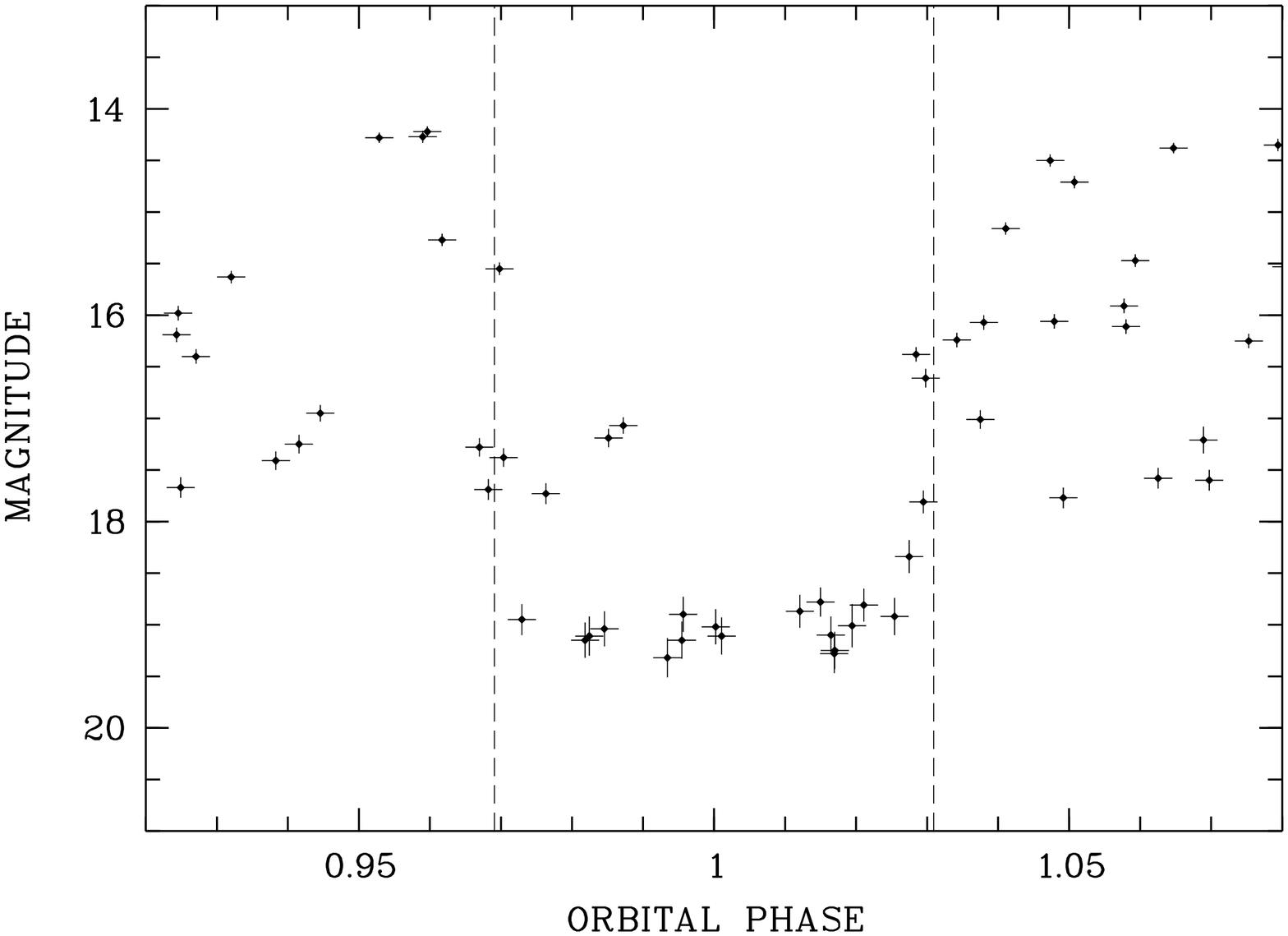}}
\caption{CRTS data of \css\ folded over orbital phase.The diagram gives 
eclipse details, the length of the horizontal bars indicate the integration time 
per measurement. The dashed vertical lines indicate the measured eclipse 
length.}
\label{f:crtsecl}
\end{figure}
%=== end figure

\subsection{The photometric behavior of \nick\ through high and low states\label{s:lcs}}
\subsubsection{CRTS photometry}
The CSS long-term light curve shown in Fig.~\ref{f:poa} covers the period 
between November 2005 and October 2013. The transient behavior was recognized 
in late 2008 but pronounced photometric variability is evident in archival data even 
prior to this.
An initial low state at $V \sim 18$ was followed by a state of intermediate brightness 
between MJD 54000 and 54200 at $V\sim16$ (September 2006 till April 2007). 
The object then returned to a low state  before turning bright with $V_{\rm max} = 14.2$ 
in late 2008. This was followed by another low state
in the fall of 2009. \nick\ was observed in intermediate to high states since then. 
Overall, the CSS-data obtained in white light show variability with an amplitude 
of about 5 mag, the maximum brightness is at 14.2, the minimum at 19.3.
A considerable number of CSS-measurements revealed the object at 19th mag, 
i.e.~about one magnitude below the average low-state brightness.

Phase-folding of all CSS data (385 data points) gives further insight (see Figs.~\ref{f:crtsfol}
and \ref{f:crtsecl}). It firstly shows that all the CSS measurements at 19th magnitude
fall into the eclipse. The eclipse light curve is flat-bottom at $m_{\rm CSS} \simeq 19$. 
This is about 1.9 mag brighter than the low-state 
V-band eclipse brightness measured by \cite{thorne+10}
at $V=20\fm86\pm0\fm05$.
During the eclipse in the low state 
%(indicated by a flat-bottom eclipse shape)
the remaining light originates mainly on the companion star. At a period of 117\,min 
one expects $V-I=2.54$ if the donor follows the CV sequence described by \citep{knigge07}. 
No color transformation into the CSS-system is available for such red stars, 
but given the red sensitivity of the CSS-system it appears reasonable that 
both eclipse brightness measurements are compatible with each other.

In the CRTS the object displays out-of-eclipse variability between bright and faint phases 
with a minimum amplitude of about 0.2\,mag. 

%=== VATT-UBV
\begin{figure}[t]
\resizebox{\hsize}{!}{\includegraphics[clip=]{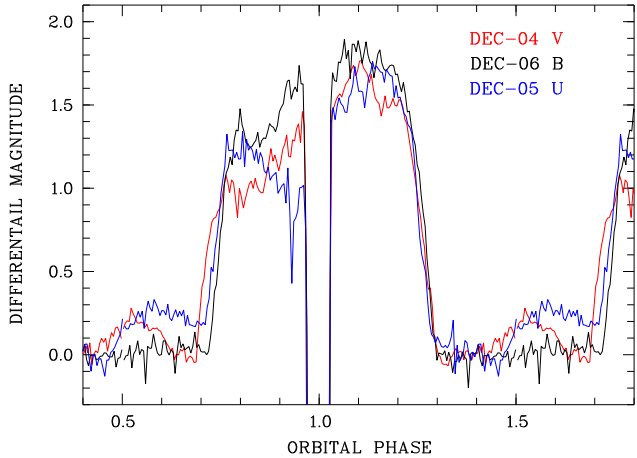}}
\caption{UBV-photometry obtained with the VATT in October 2010. Shown are
phase-averaged data (binsize 0.005 phase units, $\sim$35 seconds). Differential 
magnitudes were corrected to the same mean value around phase 0.4.}
\label{f:vatt}
\end{figure}
%=== end figure

\subsubsection{Multi-site phase-resolved photometry}
The orbital variability pattern in high and low states cannot be read from the CRTS data but only from 
phase-resolved observations. Time-resolved and phase-averaged light curves obtained by us
are shown in Figs.~\ref{f:lcoks1}, \ref{f:lcoks2}, and \ref{f:lcfol}. 
Figures \ref{f:lcoks1} and \ref{f:lcoks2} contain only data from 2009 obtained at Hankasalmi.
Phase-folded light curves from other observatories obtained between 2010 and 2014 
are shown in Fig.~\ref{f:lcfol}. 
Plotted in Figures \ref{f:lcoks1} and \ref{f:lcfol} is the flux ratio as a function of the  orbital phase. 
The data shown in the top panel were obtained in a low state \citep{thorne+10}. All other 
data in this Figure represent high accretion states. 

The light curves show commonalities and some marked differences. 
All light curves, irrespective of the accretion state, display the alternating bright-/faint-phase pattern
due to the self-eclipse of the accretion region (or regions).

The data obtained at Hankasalmi in the year 2009 can be sorted into four  distinct 
accretion states. Transitions between different states are observed to occur on timescales of days 
or even a few binary cycles. The transition from an intermediate to a high state happened within 
two binary cycles, see Fig.~\ref{f:lcoks1}.

Changes of the overall brightness are accompanied by changes of the 
start, end, and length of the bright phase and the occurrence of a pre-eclipse dip. 
The eclipse light curves are also variable and will be discussed in the next section.

The length of the bright phase changes from $\Delta \phi = \phi_{\rm end} - \phi_{\rm start} = 1.30 - 0.75 = 0.55$ in the low state 
to $1.26 - 0.63 = 0.63$ phase units in the high state (see Figs.~\ref{f:lcoks2} and \ref{f:lcfol}).
The center of the bright phase thus moves from $1.03\pm0.02$ in the low state to 
$0.98\pm0.02$ in the intermediate and finally to $0.95\pm0.02$ in the
high state, hence changing from a trailing spot at longitude $\psi \sim -10\degr$ 
to a leading spot at $\psi \sim 18\degr$. 
Some care should be taken when interpreting these numbers purely in geometrical terms, since the 
measured spot location may depend on the filter used and thus not necessarily indicate a change 
of the accretion geometry.
This becomes clear from the VATT 
UBV photometry in December 2010 (Fig.~\ref{f:vatt}).
Those data show a color-dependant start of the bright phase, whereas the end of the bright phase 
is not dependent on the filter. The accretion state is assumed to be the same.

Most of the high-state light curves show a more gradual increase of brightness at the beginning 
of the bright phase and a step decline into the self-eclipse of the accretion area. During the STELLA 
observation in 2013, however, the object displays the opposite behavior. At this epoch 
the otherwise prominent pre-eclipse dip has also vanished.
Hence, in April 2013 the accretion geometry seemed to be rather different than 
at the other occasions in the high state. 

The occurrence and the phase of the pre-eclipse dip is also dependent on the 
accretion state. The dip is typically present in the high state 
(but not during the STELLA observation in 2013) and has a typical depth of 1.4 mag.
It is centered on phase 0.86, but the phase interval $0.82-0.90$ is affected by absorption. 
In the intermediate state the dip is less pronounced and it seems to show a larger
phase jitter. The average depth is 0.3 mag but may be as large as 1.5 mag.
The phase interval $0.84-0.92$ is affected by absorption, the average dip occurs
at a later phase than during the high state, $\phi_c = 0.89$. 
No dip is detected during low and very low states.

Pre-eclipse dips are observed in several polars, a very prominent example being HU Aqr \citep{schwope+01}. 
They are caused by absorption of radiation originating from the accretion column 
in the magnetically guided part of accretion stream (or accretion curtain). The orbital phase and the extent 
of this feature gives clues to the location of the coupling region and to the 
distribution of absorbing matter in the magnetosphere of the white dwarf. 
The absence of this feature at some occasions while previously present might thus indicate 
that the matter is accreted then by another region below the 
orbital plane. 
 
The VATT V-band light curve obtained in 2009 (top panel in Fig.~\ref{f:lcfol}) is markedly 
different than the high state light curves. The rise to and the fall from the bright phase 
show a steep gradient. Also, there is a broad flux depression centered
on phase 0.1. Those properties of the light curve are reminiscent of WW Hor or DP Leo
where the light curves were shaped by strong cyclotron 
beaming \citep{beuermann+90, schwope+mengel97}. 

Occasionally a secondary maximum is observed during the faint phase
(see e.g.~the VATT V- and U-band light curves from 2010 in Fig.~\ref{f:vatt}, 
the STELLA g-band light curve from 2013 in Fig.~\ref{f:lcfol} 
and the white-light data obtained 2009, March 26, in Fig.~\ref{f:lcoks1}). 
At other  occasions the faint-phase brightness is more or less constant or just gradually increasing. 
This secondary hump has a maximum amplitude of 0.3 mag and is located at phase 
0.50 (April 2013), 0.54 (October 2010, V), and 0.59 (October 2010, U), respectively. 
Various radiation components may contribute to the observed flux: a
second accretion region, the accretion stream, the donor star, and the white dwarf.
Photospheric emission from the stars appears unlikely to us because those radiation 
components should be more prominent during low accretion states. 

\subsection{Eclipse light curves and binary parameters}
Figures~\ref{f:ecl} and \ref{f:crtsecl} show eclipse details from phase-resolved 
observations presented here or from the CSS database. 
The same phase-resolved data were used as in Fig.~\ref{f:lcfol} but with finer binning 
(0.004 phase units).
During the low state the light curve in the eclipse is flat. 
It is thus reasonable to assume that the residual emission is due to the 
donor star \citep{thorne+10}. During intermediate and high states rather  
intense emission is remaining throughout the eclipse of the white dwarf. 
This is due to the accretion stream/curtain outside the orbital plane.
From its initial brightness after eclipse ingress it fades typically by two magnitudes but 
seems to stay above the brightness of the donor star, hence some parts 
of the stream are always visible.

Most of the CRTS data that were obtained the white-dwarf eclipse 
show the source at $V_{\rm CSS} = 19$ but a few data points showed the object 
considerably brighter, $17.0 - 17.7$\,mag. 
These data points were obtained during very high accretion states and the high 
brightness is also due to the bright accretion stream.

%===
\begin{figure}[t]
\resizebox{\hsize}{!}{\includegraphics[clip=]{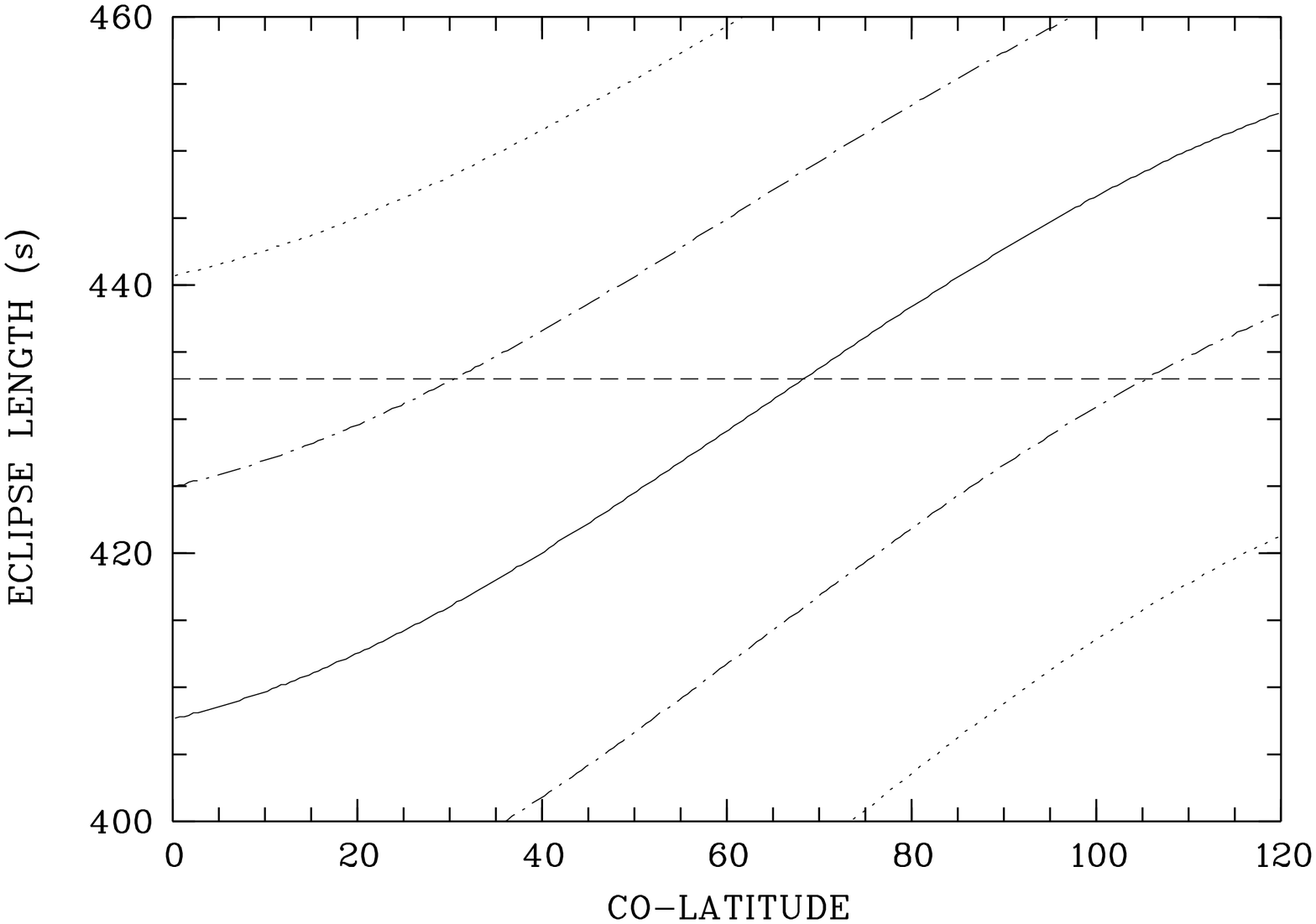}}
\caption{Eclipse length of points along the central meridian for an 0.8\,\msun\ white 
dwarf as a function of the co-latitude $\beta$ and the orbital inclination.
The measured eclipse length of 433.08\,s is illustrated by the dashed horizontal
line. 
}
\label{f:eclele}
\end{figure}
%=== end figure

The eclipse width $\Delta\phi_{\rm ecl}$, the orbital inclination $i$ and the mass ratio $Q$ 
are related to each other via the Roche geometry \citep{chanan+76}. Taking the observed 
eclipse width at face value, i.e.~as a proxy for the eclipse width of the white dwarf, 
a relation between between $i$ and $Q$ can be derived. If one further assumes
that the donor star follows Knigge's sequence,
$M_2 = 0.16$\,\msun\ \citep{knigge07}, and assuming 
$M_1 = 0.8$\,\msun, a typical value for the white dwarfs in CVs, the 
orbital inclination is $i=81.7\degr$. For a standard single white dwarf, $M_1 = 0.6$\,\msun,
the inclination would be $79.8\degr$ and for a more massive white dwarf with 
$M_1 = 1.0$\,\msun\ one gets $i = 83.2\degr$. A further uncertainty derives from the 
fact, that the observed eclipse width does not exactly correspond to the eclipse 
of the white dwarf centre of mass. It belongs mainly to the accretion spot somewhere on the
surface of the white dwarf. In order to estimate the size of this effect, we have calculated 
eclipse light curves for certain points on a sphere. A white dwarf with
0.8\,\msun\ was assumed and the points were chosen along the central meridian (meridian 
on the white dwarf through the rotation axis and the line connecting both stars). 
Eclipse light curves were computed as a function of the stellar co-latitude $\beta$, i.e.~the
angle between the rotation axis and the point on the meridian. Results are shown 
in Fig.~\ref{f:eclele} for $i=81.7\degr, \pm 0.5\degr,\pm1.0\degr$, respectively. 
The measured eclipse length is the same for an accretion spot at $\beta = 68\degr$ and 
at the center of mass of the assumed 0.8\,\msun white dwarf. The computations show that 
one should consider another $\sim$0.5\degr\ systematic uncertainty of the inclination
derived straight from the observed eclipse width via Roche-lobe geometry. 
The inclination is thus expected to be in the range $i=79.3\degr - 83.7\degr$.

\section{Discussion and conclusion\label{s:con}}
The present study presents an analysis of multi-epoch time-resolved 
photometric observations of the bright eclipsing polar \css. We have established a 
precise long-term ephemeris based on 109 individual timings of the eclipse obtained 
between 2009 and 2014. We also determined a precise eclipse length of $433.08\pm0.65$\,s.

An eclipse length of 
$408 \pm7$\,s reported by \citet{thorne+10} can definitely be ruled out already on the basis of a few light curves 
that were obtained with sufficient time resolution. The measured eclipse length was used 
to constrain the orbital inclination to lie within the range $i=79.3\degr - 83.7\degr$. The uncertainty 
is due to the unknown, hence assumed mass of the white dwarf and the unknown location 
of the accretion spot. Photometric observations with high time resolution during a 
low state would be most useful to directly measure the size of the white dwarf and
high-speed photometric observations during a high state would allow 
to measure the size and location (azimuth) of the accretion spot directly. 

We find a linear relation between cycle number and mid-eclipse time which is stable 
over a time interval of 5.3 years or almost 24000 orbital cycles of the binary. We estimate 
the maximum mass of an unseen companion by adding one higher polynomial order
and explore the $\chi^2$ space until the fit deteriorates. We request a 95\% confidence
to reject the hypotheses of a quadratic term in the ephemeris. The bow of the parabola 
is used as a proxy for the amplitude of the sine curve due to the LTT effect. 
A period 5.3 years is assumed for the putative substellar companion. Its amplitude 
then is about 3 sec (6 s peak-to-peak) which corresponds to $M_{\rm pl} \simeq 2$\,M$_{\rm Jup} $
for a negligible eccentricity, for an edge-on planetary system and for a total mass of the 
system of 1\,\msun. 
Care has to be taken to not overinterpret this result. It may well be that the system 
is on an apparently linear part of the $O-C$ diagram over the past 5 years which over a longer 
time base might become saw-tooth shaped \citep[for an example see NN Ser, ][]{beuermann+10}. 
Monitoring of the system for several years to come is needed to confirm or disprove 
the linear ephemeris. The current work defines the reference for such future studies.

Monitoring over long times is also needed to measure the center of the bright phase 
and probe the coupling mechanism and stability of the white dwarf orientation in the system. 
With a monitoring campaign lasting more than 30 years \citet{beuermann+14} have 
uncovered the likely oscillation of the magnetic axis around an equilibrium position 
of the white dwarf in the eclipsing polar DP Leo. The spot longitude of DP Leo 
was observed at $-10\degr$ to $-15\degr$ some 30 years ago, similar to what is observed
in \nick. Contrary to DP Leo the longitude of the accretion spot in \nick\ 
is observed to vary as a function of the mass accretion rate. 
In the low state the spot lies at $-10\degr$, and in the high state at $+10\degr$. 
The accretion-rate dependent location of the accretion spot 
complicates the proposed investigation. The high brightness 
of \nick\ on the other hand is promising to establish a large database and to separate
oscillations of the magnetic axis and spot migrations.

\begin{figure*}[t]
\resizebox{0.49\hsize}{!}{\includegraphics[clip=]{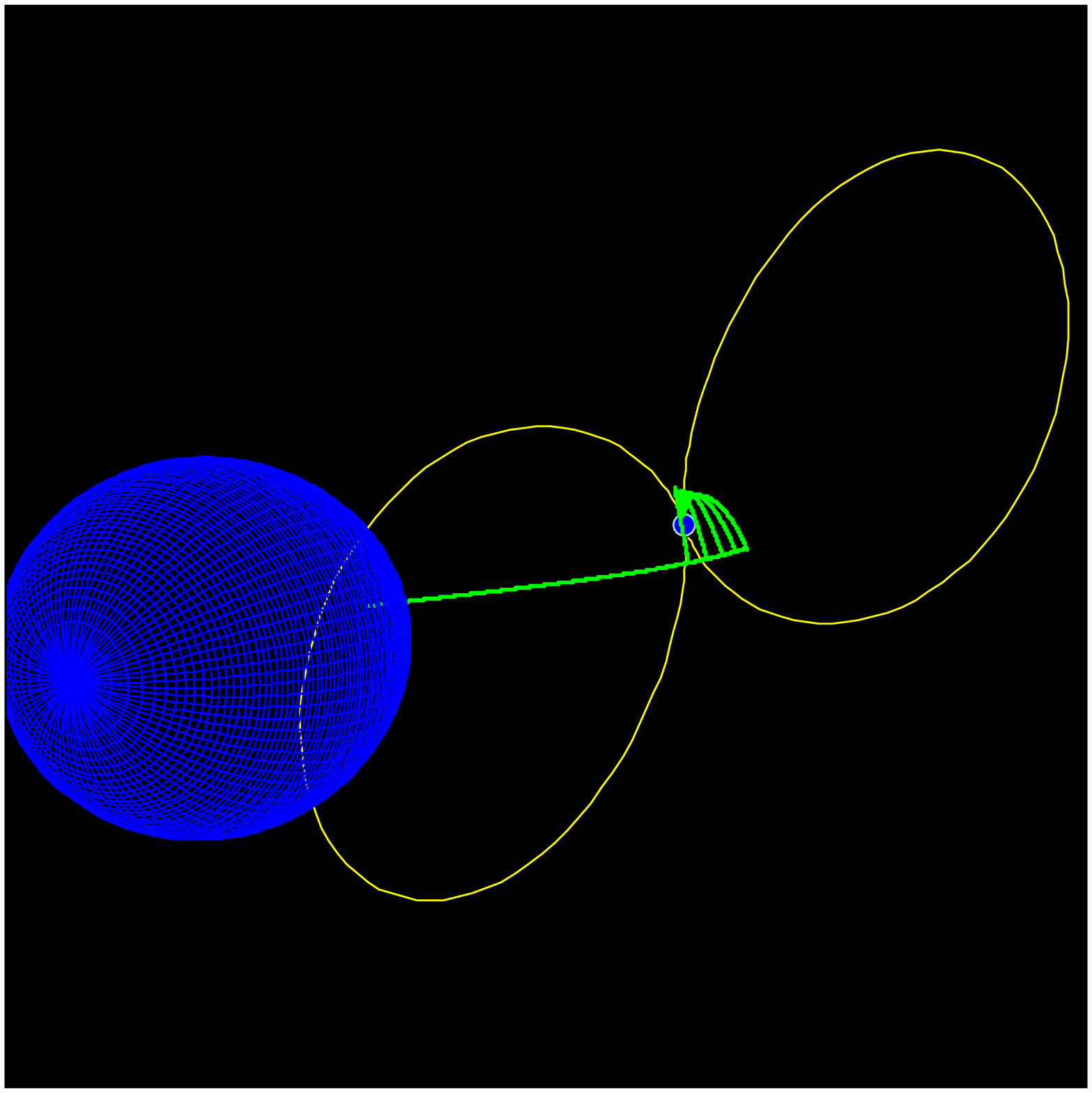}}
\resizebox{0.49\hsize}{!}{\includegraphics[clip=]{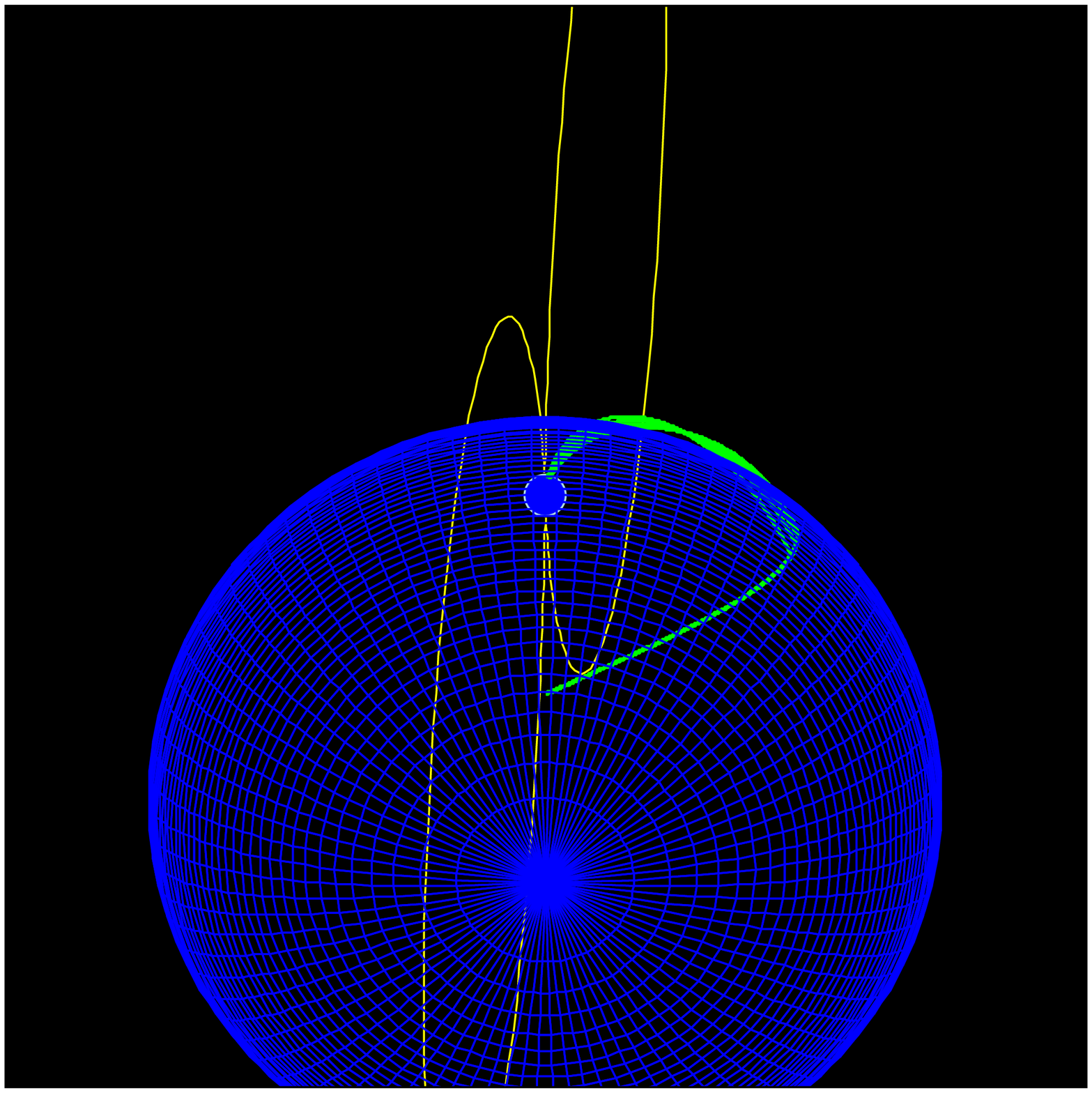}}
\caption{Possible accretion geometry of \css\ in a high state 
seen at binary phases 0.9 and 1.0. The green lines illustrate accreting field 
lines, the yellow lines are dipolar field lines through the magnetic meridian thus illustrating
the orientation of the magnetic field. The orbital inclination is 79.3\degr, the mass ratio $Q=5$ and the 
mass of the white dwarf $M_{\rm WD} = 0.8$\,\msun.}
\label{f:view1}
\end{figure*}

The final part of the paper is devoted to a discussion of the possible accretion geometry.
An accretion arc is thought to arise at the foot points
of dipolar field lines connecting the ballistic accretion stream with the white dwarf. 
Possible solutions are sought with the following constraints: -- the accretion spot is 
centered on phase 0.95 in the high state and 1.03 in the low state. Thus the spot
longitude is $\psi = +18\degr$ in the high state and $\psi = -10\degr$ in the low state 
(measured with respect to the line connecting both stars in a
coordinate system centered on the white dwarf);
-- the inclination is between $i=79.3\degr$ and 83.7\degr;
-- in the high state parts of the accretion stream are never completely obscured by the donor star;
-- the stagnation region is extending from about 35\degr\ to about 65\degr\ in azimuth in the 
high state, and from $\sim$30\degr\ to $\sim$60\degr\ in the intermediate state 
(phase interval of pre-eclipse dip).
The absence of pre-eclipse dips in the low state prevents us 
from constraining the stagnation region when the accretion rate is further reduced. 

Firstly we are discussing a possible high-state accretion geometry.
The presence of a pre-eclipse dip requires an accretion region above the orbital plane.
An only partial occultation of the accretion stream requires 
a sufficiently low orbital inclination and a sufficiently high elevation of the stream. 
The latter becomes possible if the inclination of the magnetic axis is sufficiently small.
A location of the accretion region closer to the line connecting both stars than the 
stagnation region means that the magnetically guided part of the stream has to be
strongly re-directed in the frame of the binary.

We use our own visualization software to test different accretion scenarios.
The original free-falling stream is eventually guided by magnetic field lines towards 
an accretion region on the white dwarf. The coupling region is found by balancing 
the ram and the magnetic pressure along the stream. In the coupling region 
the stream is assumed be instantaneously re-directed and traveling along dipolar
field lines. We find a possible accretion geometry for the high state assuming
a co-latitude and an azimuth of the magnetic axis of $\beta=18\degr$, and $\psi = -3\degr$, and
an inclination of 79.3\degr.
For the chosen orientation of the magnetic axis the accretion stream would be completely 
occulted at phase zero if the orbital inclination were higher. A higher inclination would be 
possible if the magnetic axis were more closely aligned to the rotation axis. In such a geometry 
the magnetically guided part cannot be re-directed to feed an accretion region close 
to the line connecting both stars. A sketch of the accretion geometry is 
given in Fig.~\ref{f:view1} where the binary is seen at orbital phases 0.9, i.e.at the end
of the pre-eclipse dip, and at phase 0.0.

At a lower accretion rate coupling occurs earlier and the magnetically guided part of the 
stream is lifted further above the plane. Hence, at lower accretion rate the magnetically guided
part is not expected to be completely covered by the donor during eclipse phase.
The pre-eclipse dip occurs later in phase, 
i.e.~closer to the eclipse, and consequently the accretion region lies closer to the 
meridian. All those expectations are met in the intermediate state. 
Hence, the high and intermediate states can be 
explained with the same accretion geometry. 

The same arguments apply if the accretion rate is further reduced, i.e.~if one wants to 
understand the geometry in the low state. One might expect to always see some remaining 
radiation from the magnetic stream lurking at high elevations above the orbital plane.
Contrary to this a flat eclipse light curve  is observed suggesting that matter in the magnetosphere 
is completely covered by the donor star during eclipse in the low state.
Also, the observed location of the accretion spot, $\psi_{\rm spot} \simeq -10\degr$
is incompatible with the chosen orientation of the magnetic axis, $\psi = -3\degr$.
If a more extreme value of $\psi$ is chosen, $\psi < 10\degr$, the high-state geometry cannot 
be explained. 

In short, we do not find a solution which 
satisfies all observational constraints during high and low accretion states. 
A possible remedy might be found in assuming a non-dipolar field structure. A further 
variation of the theme would be assuming that accretion takes place at a region 
below the orbital plane in the low state. This would still require a non-dipolar field structure but 
would explain the flat eclipse light curves. The absence of the pre-eclipse dip 
in the STELLA light curves obtained in 2013 remains puzzling.

\acknowledgement
We thank an anonymous referee for helpful comments.
 
We thank Gerhard Dangl for obtaining and providing his photometric data. 

This study is based partly on data obtained with the STELLA robotic telescope in Tenerife, an AIP facility 
jointly operated by AIP and IAC.

We thank the Krizmanich family for their generous donation 
of the Sarah L.~Krizmanich telescope to the University of Notre Dame.

Based on observations with the VATT: the Alice P.~Lennon Telescope and the 
Thomas J.~Bannan Astrophysics Facility. 
We thank Richard Boyle and the Vatican Observatory for scheduling the VATT observations.

The CSS survey is funded by the National Aeronautics and Space
Administration under Grant No. NNG05GF22G issued through the Science
Mission Directorate Near-Earth Objects Observations Program.  The CRTS
survey is supported by the U.S.~National Science Foundation under
grants AST-0909182 and AST-1313422.

\bibliographystyle{aa}
\bibliography{css}

\onecolumn
\begin{longtable}{r r@{.}l c c}
\caption{Cycle number and mid-eclipse times of \nick\ \label{t:ecl}}\\ %\hline
Cycle  & \multicolumn{2}{c}{$ T_C $ in BJD(TDB)} &  $\Delta T_C$ & Observatory$^{\rm a)}$ \\ \hline
\endfirsthead
\caption{Cycle numbers and mid-eclipse times of \nick\ (continued)} \\ \hline
Cycle  & \multicolumn{2}{c}{$ T_C $ in BJD(TDB)} &  $\Delta T_C$ & Observatory$^{\rm a)}$ \\ \hline
\endhead
\hline
\endfoot 
\hline
\endlastfoot
    0 & 2454833 & 20785 & 1.6E-4 & 1\\
    1 & 2454833 & 28923 & 1.6E-4 & 1\\
    2 & 2454833 & 37055 & 1.6E-4 & 1\\
    3 & 2454833 & 45186 & 2.2E-4 & 1\\
   16 & 2454834 & 50993 & 1.6E-4 & 1\\
   17 & 2454834 & 59129 & 1.6E-4 & 1\\
   18 & 2454834 & 67258 & 2.2E-4 & 1\\
   40 & 2454836 & 46319 & 2.2E-4 & 1\\
   41 & 2454836 & 54428 & 2.2E-4 & 1\\
   74 & 2454839 & 22969 & 1.6E-4 & 1\\
   75 & 2454839 & 31113 & 1.6E-4 & 1\\
   76 & 2454839 & 39258 & 1.6E-4 & 1\\
  948 & 2454910 & 35315 & 1.6E-4 & 1\\
  949 & 2454910 & 43440 & 1.6E-4 & 1\\
  973 & 2454912 & 38734 & 1.6E-4 & 1\\
  974 & 2454912 & 46894 & 1.6E-4 & 1\\
 1009 & 2454915 & 31692 & 1.6E-4 & 1\\
 1010 & 2454915 & 39844 & 1.6E-4 & 1\\
 1022 & 2454916 & 37481 & 1.6E-4 & 1\\
 1023 & 2454916 & 45626 & 1.6E-4 & 1\\
 1024 & 2454916 & 53780 & 1.6E-4 & 1\\
 1034 & 2454917 & 35159 & 1.6E-4 & 1\\
 1035 & 2454917 & 43294 & 1.6E-4 & 1\\
 1036 & 2454917 & 51423 & 1.6E-4 & 1\\
 1037 & 2454917 & 59568 & 1.6E-4 & 1\\
 1046 & 2454918 & 32825 & 1.6E-4 & 1\\
 1048 & 2454918 & 49069 & 1.6E-4 & 1\\
 1049 & 2454918 & 57211 & 1.6E-4 & 1\\
 1107 & 2454923 & 29200 & 1.6E-4 & 1\\
 1108 & 2454923 & 37340 & 1.6E-4 & 1\\
 1182 & 2454929 & 39531 & 3.0E-5 & 2\\
 1194 & 2454930 & 37167 & 1.6E-4 & 1\\
 1206 & 2454931 & 34834 & 3.0E-5 & 2\\
 1267 & 2454936 & 31234 & 1.6E-4 & 1\\
 1280 & 2454937 & 37036 & 1.6E-4 & 1\\
 1281 & 2454937 & 45159 & 1.6E-4 & 1\\
 1282 & 2454937 & 53293 & 1.6E-4 & 1\\
 1316 & 2454940 & 29976 & 1.6E-4 & 1\\
 1317 & 2454940 & 38110 & 1.6E-4 & 1\\
 1318 & 2454940 & 46269 & 1.6E-4 & 1\\
 1319 & 2454940 & 54390 & 1.6E-4 & 1\\
 1366 & 2454944 & 36857 & 1.6E-4 & 1\\
 1367 & 2454944 & 45013 & 2.2E-4 & 1\\
 1380 & 2454945 & 50773 & 1.6E-4 & 1\\
 1391 & 2454946 & 40299 & 1.6E-4 & 1\\
 3609 & 2455126 & 89667 & 1.2E-4 & 3\\
 3610 & 2455126 & 97805 & 1.2E-4 & 3\\
 3621 & 2455127 & 87319 & 1.2E-4 & 3\\
 3622 & 2455127 & 95460 & 1.2E-4 & 3\\
 3634 & 2455128 & 93112 & 8.2E-5 & 3\\
 3635 & 2455129 & 01256 & 8.2E-5 & 3\\
 3646 & 2455129 & 90767 & 1.2E-4 & 3\\
 4980 & 2455238 & 46428 & 1.6E-4 & 1\\
 4981 & 2455238 & 54570 & 1.6E-4 & 1\\
 4982 & 2455238 & 62707 & 1.6E-4 & 1\\
 5102 & 2455248 & 39235 & 1.6E-4 & 1\\
 8621 & 2455534 & 75729 & 4.1E-5 & 3\\
 8622 & 2455534 & 83859 & 6.2E-5 & 3\\
 8623 & 2455534 & 92003 & 6.2E-5 & 3\\
 8624 & 2455535 & 00136 & 1.2E-4 & 3\\
 8634 & 2455535 & 81515 & 1.2E-4 & 3\\
 8635 & 2455535 & 89666 & 1.2E-4 & 3\\
 8636 & 2455535 & 97797 & 1.2E-4 & 3\\
 8646 & 2455536 & 79164 & 1.2E-4 & 3\\
 8647 & 2455536 & 87304 & 1.2E-4 & 3\\
 8648 & 2455536 & 95449 & 1.2E-4 & 3\\
10226 & 2455665 & 36704 & 1.6E-4 & 1\\
10227 & 2455665 & 44848 & 1.6E-4 & 1\\
10239 & 2455666 & 42488 & 1.6E-4 & 1\\
10240 & 2455666 & 50654 & 1.6E-4 & 1\\
10300 & 2455671 & 38878 & 1.6E-4 & 1\\
10301 & 2455671 & 47052 & 1.6E-4 & 1\\
17065 & 2456221 & 90311 & 3.7E-5 & 5\\
17075 & 2456222 & 71689 & 3.7E-5 & 5\\
17076 & 2456222 & 79827 & 3.7E-5 & 5\\
17112 & 2456225 & 72787 & 3.7E-5 & 5\\
17113 & 2456225 & 80928 & 3.7E-5 & 5\\
17224 & 2456234 & 84201 & 3.7E-5 & 5\\
17225 & 2456234 & 92344 & 5.3E-5 & 5\\
17259 & 2456237 & 69022 & 3.7E-5 & 5\\
17260 & 2456237 & 77157 & 3.7E-5 & 5\\
17261 & 2456237 & 85298 & 3.7E-5 & 5\\
17381 & 2456247 & 61809 & 6.2E-5 & 4\\
17382 & 2456247 & 69949 & 6.2E-5 & 4\\
17418 & 2456250 & 62909 & 6.1E-5 & 4\\
18390 & 2456329 & 72733 & 8.2E-5 & 3\\
18391 & 2456329 & 80863 & 4.1E-5 & 3\\
19062 & 2456384 & 41264 & 2.5E-4 & 4\\
19136 & 2456390 & 43435 & 2.5E-4 & 4\\
19172 & 2456393 & 36392 & 2.5E-4 & 4\\
19173 & 2456393 & 44550 & 2.5E-4 & 4\\
19197 & 2456395 & 39847 & 2.5E-4 & 4\\
19198 & 2456395 & 47970 & 2.5E-4 & 4\\
23499 & 2456745 & 48149 & 1.3E-5 & 6\\
23500 & 2456745 & 56289 & 1.3E-5 & 6\\
23510 & 2456746 & 37666 & 1.3E-5 & 6\\
23511 & 2456746 & 45803 & 1.3E-5 & 6\\
23513 & 2456746 & 62084 & 1.2E-4 & 5\\
23514 & 2456746 & 70220 & 1.3E-4 & 5\\
23522 & 2456747 & 35317 & 1.3E-5 & 6\\
23525 & 2456747 & 59729 & 4.5E-5 & 5\\
23526 & 2456747 & 67867 & 4.5E-5 & 5\\
23599 & 2456753 & 61920 & 1.9E-5 & 7\\
23599 & 2456753 & 61921 & 4.5E-5 & 5\\
23600 & 2456753 & 70053 & 1.3E-5 & 7\\
23600 & 2456753 & 70060 & 4.5E-5 & 5\\
23636 & 2456756 & 63019 & 1.3E-5 & 7\\
23707 & 2456762 & 40788 & 1.3E-5 & 6\\
23722 & 2456763 & 62851 & 4.5E-5 & 5\\
\end{longtable}
$^{\rm a)}$ Observatory key: 1 -- Hankasalmi, 2 -- Nonndorf, 3 -- VATT, 4 -- STELLA , 5 -- SCT28, 6 -- IOP, 7 -- SLKT 
\twocolumn

\end{document}